# Formation mechanisms of single-crystalline InN quantum dots fabricated via droplet epitaxy

P. Aseev[a,b], Ž. Gačević[a,*], J.M. Mánuel[c,d], J.J. Jiménez[c,d], R. García[c,d], F.M. Morales[c,d], and E. Calleja[a]

[a] *Instituto de Sistemas Optoelectrónicos y Microtecnología, Universidad Politécnica de Madrid, Ciudad Universitaria s/n, 28040 Madrid, Spain*

[b] *QuTech and Kavli Institute of Nanoscience, Delft University of Technology, 2628 CJ Delft, The Netherlands*

[c] *IMEYMAT: Institute of Research on Electron Microscopy and Materials of the University of Cádiz, Puerto Real, 11510 Cádiz, Spain*

[d] *Department of Materials Science and Metallurgic Engineering, and Inorganic Chemistry, Faculty of Sciences, University of Cádiz, Puerto Real, 11510 Cádiz, Spain*

This work presents an experimental and theoretical insight into formation mechanisms of single crystalline wurtzite InN quantum dots (QDs) fabricated via metal droplet epitaxy (DE) by employing plasma assisted molecular beam epitaxy. The applied procedure consists of two fabrication stages. During the first stage, the cold substrate ($T \approx 15$ °C) is exposed to an impinging In flux, resulting in formation of metallic In droplets on the substrate surface, and then to an impinging active nitrogen flux, resulting in In conversion into polycrystalline InN islands. During the second stage, the substrate, which is still kept exposed to active nitrogen, is heated up to $T \approx 300$ °C, to allow for the reorganization of extended polycrystalline InN islands into groups of independent single-crystalline wurtzite InN QDs. This work provides a detailed experimental insight into both fabrication stages and their qualitative explanations within the scopes of adatom surface kinetics (stage I) and total energy per unit crystal volume minimization (stage II). Finally, the formation mechanisms of InN QDs on the three different substrates (Si(111), Si(001) and $In_{0.3}Ga_{0.7}N/Si(111)$) are compared, and also linked to the formation mechanisms of other more studied nanostructures, such as self-assembled GaN/AlN QDs and self-assembled and selective-area-grown GaN nanowires.

---

[*] Corresponding author. E-mail address: gacevic@isom.upm.es (Ž. Gačević).

# 1. Introduction

III-Nitride quantum dots (QDs) structures are becoming increasingly important as active regions of various (opto)electronic devices, such as light emitting and laser diodes [1], single photon sources [2], photodetectors [3], solar cells [4,5] as well as various chemical- and bio-sensors [6,7]. In the specific case of InN QDs, recently, a very promising application has emerged in solar energy-assisted water-splitting devices, where it has been demonstrated that decorating an $In_xGa_{1-x}N$ photoanode with surface InN QDs can nearly double the device efficiency (incident photon to device current conversion efficiency) [8]. While new engineering proposals concerning III-Nitride QDs are constantly emerging, the works reporting new strategies for their synthesis, improved fabrication controllability and/or better understanding of their formation mechanisms, remain very scarce. Well-controlled and precisely understood methods for III-Nitride QDs fabrication thus still remain a huge challenge.

Over the last two decades many works reported growth of $In_xGa_{1-x}N$ QDs, focusing mainly on their potential to improve efficiency of classical light emitters (light emitting diodes and laser diodes) and to act as an active region of quantum light emitters (single photon sources). In the case of former structures, which usually have planar geometry, the QDs are most commonly grown via Stranski-Krastanov method [9,10]; in the case of latter structures, which usually have columnar geometry, the QDs are most commonly formed via spontaneous In clustering within the $In_xGa_{1-x}N$ ternary [11]. Note that in the case of the Stranski-Krastanov method, it is strain accumulated in a thin $In_xGa_{1-x}N$ epitaxial film, which act as a driving mechanism for QD formation; in the case of spontaneous In clustering, it is a large miscibility gap of $In_xGa_{1-x}N$ ternary which drives QD formation [12]. To undergo from 2D to 3D morphological transition, a certain amount of strain must accumulate in the $In_xGa_{1-x}N$ epilayer, i.e. its composition must fall in a specific range and the layer itself must be grown on a flat underlying substrate. Similarly, In clustering is effective for formation of $In_xGa_{1-x}N$ QDs only for a specific In content range (low-to-moderate); in addition, the method is spontaneous, i.e. virtually uncontrollable. As can be noticed, neither of the two most common methods is convenient for fabrication of $In_xGa_{1-x}N$ QDs with high In content. The methods are particularly inconvenient for the realization of surface In(Ga)N QDs on layers with rough morphology.

The previous limitations can be overcome if droplet epitaxy (DE) method is employed [13]. In this method, which has been widely reported for the realization of different types of III-V QDs, group III and group V elements are not supplied simultaneously, but with a certain time delay. This fabrication approach allows In adatoms to form 3D metallic droplets on the substrate surface, first, and only then to be converted into semiconductor islands, under group V element flux. Keeping the adatoms surface mobility low (via keeping the substrate "cold"), is key to form small In metallic droplets in the initial fabrication stage, and small 3D InN islands, upon the In-to-InN conversion [14]. While the current understanding of DE predicts this method to be feasible, for QD fabrication on both atomically flat and rough underlying substrates, InN QDs fabrication has been reported exclusively on flat substrates (rms ≈ 1 nm), so far, such as on: GaN/sapphire (0001) [15], AlN/Si (111) [16], and Si(111) [17-19]. In addition, no comprehensive studies concerning QD formation mechanisms have been reported.

Hereby, we report fabrication of single crystalline InN QDs on both atomically flat Si ((111) and (001)) substrates and relatively rough (rms ≈ 10 nm) $In_{0.3}Ga_{0.7}N$/Si(111) templates. Making use of scanning electron and atomic force microscopies as well as transmission and scanning-transmission electron microscopies (SEM, AFM, TEM and STEM), we provide a detailed experimental evidence concerning different stages of QD formation. The obtained results are qualitatively explained within the scopes of adatom surface kinetics and total energy per unit crystal volume minimization. Finally, formation mechanisms of InN QDs are compared with those of more studied nanostructures, such as self-assembled GaN/AlN QDs, as well as self-assembled and selective-area-grown GaN nanowires.

## 2. Experimental

The samples were grown in a MECA 2000 molecular beam epitaxy (MBE) system equipped with a radio-frequency plasma nitrogen source and standard Knudsen cells for In and Ga. Commercial Si (111) and Si (001) substrates were used as substrates. The experiments were conducted on three different substrates, atomically flat Si (111) and Si (001) substrates, as well as on relatively rough (rms ≈ 10 nm) $In_{0.3}Ga_{0.7}N$/Si(111) templates ($In_{0.3}Ga_{0.7}N$ thickness ≈ 300 nm), grown in the same MBE reactor. Both Si (111) and Si (001) were outgassed in the growth chamber at ≈ 900 °C for 30 min, to remove substrates' native oxide and provide clean Si surface. The growth temperature was measured using a thermocouple, installed close to the substrate backside. Impinging metal fluxes of In and Ga were measured *in-situ* by a flux gage in equivalent pressure units (Torr). The fluxes were calibrated in absolute units (atoms/cm$^2$·s) making use of thicknesses of compact InN and GaN layers (grown under N-rich conditions), which were measured *ex-situ* by SEM. Similarly, the active nitrogen flux was calibrated making use of thickness of compact GaN layers (grown under Ga-rich conditions) and *ex-situ* measured by SEM. The sample surface, i.e. morphology of In metallic droplets, InN polycrystalline islands and InN monocrystalline QDs, were assessed *ex-situ* by Digital Instruments MMAFM-2 AFM and FEI Inspect F50 SEM. To get precise information about InN QD morphological, structural and chemical properties a STEM JEOL2010-FEG, working at 200 kV, was utilized to carry out different techniques. These techniques included: bright field TEM (BF-TEM) micrographs, phase contrast TEM mode, also known as high resolution TEM (HRTEM), high angle annular dark field (HAADF) images and X-ray energy dispersive (EDX) spectra. Samples under study were prepared in cross-section disposition and thinned until the electron-transparency has been reached, first through a traditional mechanical method and then through posterior ion-milling (for the latter, Ar+ ions accelerated at 4 kV in a GATAN precision ion polishing system have been used). Concerning HAADF-TEM, note that intensity in images obtained in this mode is proportional to ~$t·Z·n$, where $t$ is the thickness of the material along the electron beam direction, $Z$ is the average atomic mass of the materials, and $n$ is a number, usually between 1.6 and 1.9. This implies that an intensity contrast in HAADF images can be interpreted as a change in composition and/or thickness, between different structures or phases. Local EDX spectra, taken over 5 × 5 nm$^2$, were taken in order to get local chemical composition of the studied material. To obtain EDX-maps, a FEI Titan Cubed Themis 60-300 STEM microscope, working at 200 kV, was utilized. This microscope is equipped with a Super-X windowless EDX detector (enabling four-quadrant SD EDX detection in a solid angle higher than 0.7 srad).

# 3. Results and discussion

## 3.1. Model for the droplet epitaxy of InN QDs

The DE of InN QDs can be divided into two fabrication stages (Fig. 1). The first fabrication stage (I), most likely driven by adatom surface kinetics, consists of "cold" substrate ($T_S \approx 15$ °C) exposure first to (i) In flux, resulting in formation of small hemispherical metallic In droplets with high surface density, and then to (ii) active N flux, resulting in metallic In conversion into semiconductor InN and formation of extended polycrystalline InN islands (Fig. 1, stage I). The second fabrication stage (II), most-likely driven by total-energy-per-unit-volume-minimization, consists of thermal annealing (at $T_S \approx 300$ °C) of the newly formed InN polycrystalline islands under active N flux, resulting in polycrystalline islands reordering into groups of single crystal InN QDs (Fig. 1, stage II).

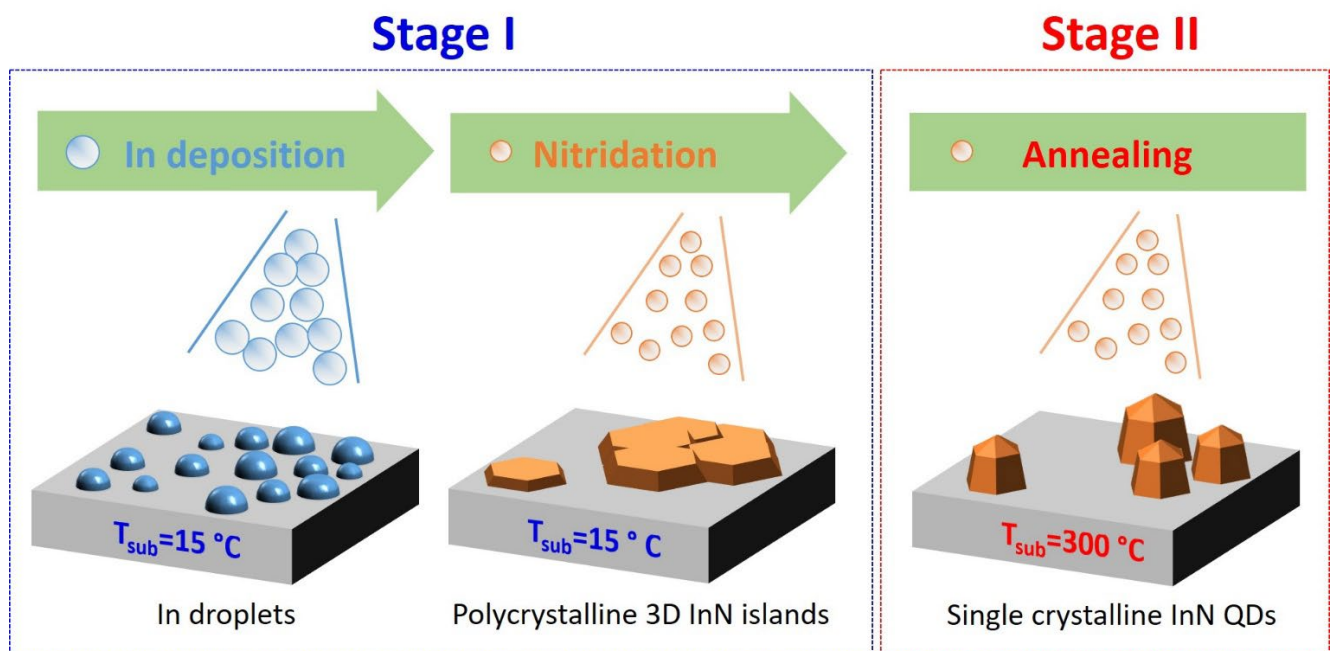

**Fig. 1.** Depiction of the steps for InN QDs fabrication via metal droplet epitaxy.

### 3.1.1 Experimental evidence

To get a detailed insight into InN QDs formation mechanisms, many samples have been grown by DE on atomically flat Si (111) substrates, applying a variety of growth conditions. To gather experimental evidence, the fabrication process was stopped after each of the three fabrication steps (In deposition, nitridation and annealing). The samples were then inspected by SEM, AFM and/or (S)TEM. In this section, the most relevant experimental findings are summarized.

**Stage I: In deposition.** The In deposition is defined by three parameters: impinging In flux ($\Phi_{In}$), In deposition time ($t_{In}$) and substrate temperature ($T_S$). In this study the first two parameters were kept constant: $\Phi_{In} \approx 3.2 \cdot 10^{14}$ atoms/cm$^2$·s (supplied by keeping the In evaporation cell at 770 °C) and $t_{In} = 10$ s, whereas the third parameter was varied within the $T_S = -10 - 110$ °C range, keeping it significantly below the In melting temperature $T_M = 156.6$ °C. The substrate temperature has no significant effect on the In droplets density/size distribution as long as it is kept below the In melting point. Fig. 2a presents an AFM scan of a sample where In deposition was performed at $T_S \approx 15$ °C. The AFM reveals formation of small hemispherical In droplets, with a surface density of $\sigma \approx 1.0 \cdot 10^{11}$ cm$^{-2}$, average height of $H = 1.5$ nm ($\sigma_H = 1.2$ nm) and diameter of $D = 15.0$ nm ($\sigma_D = $ 1.2 nm). We note here that due to the finite size of the AFM "needle" tip radius, typically ≈10 nm, the AFM can significantly overestimate the diameter of such small and closely-packed objects, thus we must be very cautious with interpretation of AFM results. Several samples with substrate temperatures above the In melting point ($T_S > 156.6°C$) were also fabricated, for the study completeness. In that case, In droplets with much lower surface density, typically $\sigma < 1.0 \cdot 10^5$ cm$^{-2}$, and exceedingly large size, typical diameters in $D = 1 - 10$ μm range, were formed (results not presented). Further explanations considering the mechanisms which drive this fabrication step are given hereafter.

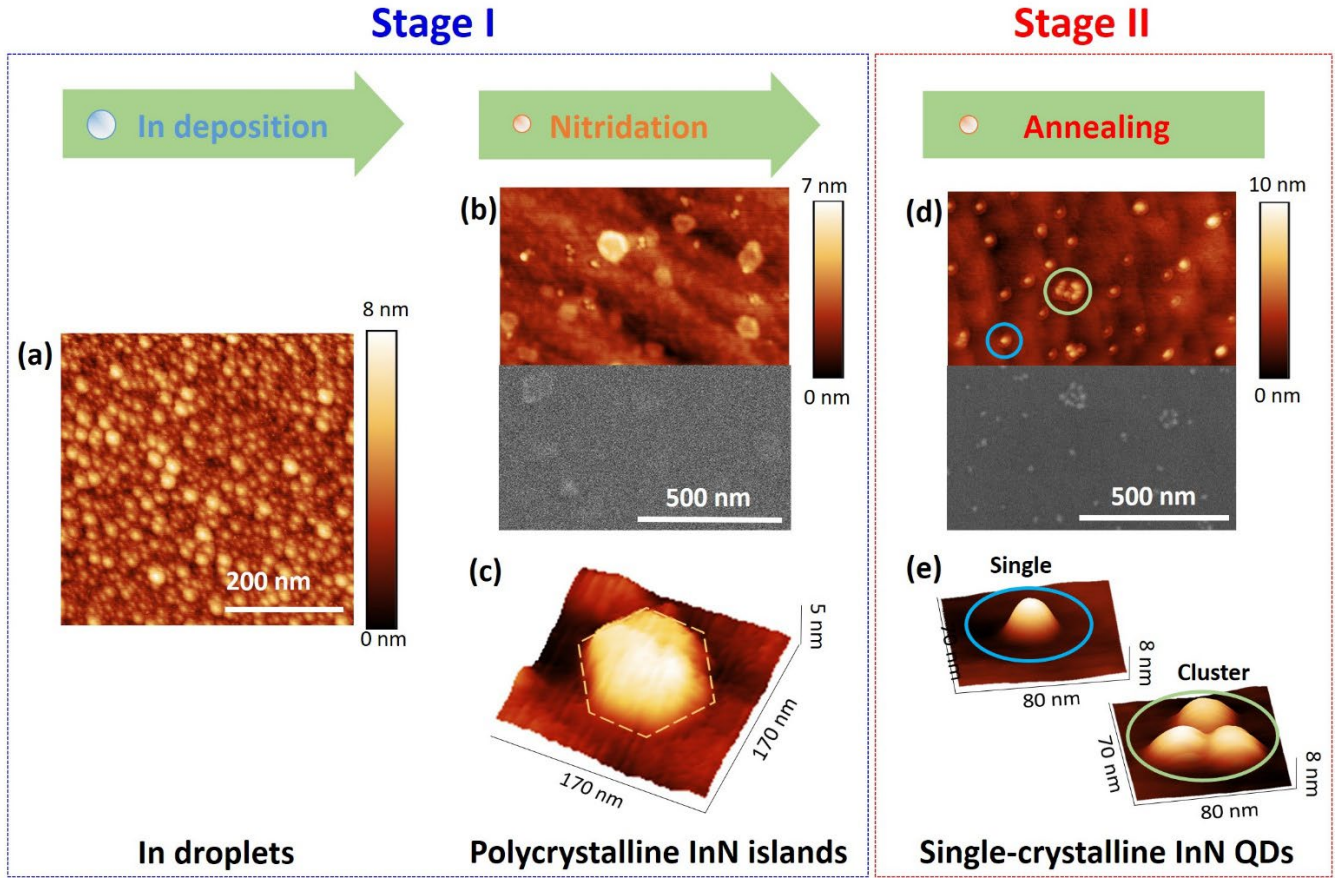

**Fig. 2.** Morphological properties of (a) In droplets, (b-c) polycrystalline InN islands and (d-e) single crystal InN QDs, as assessed by AFM (colored) and SEM (grayscale).

**Stage I: Nitridation.** The nitridation step is also defined by three parameters: impinging N flux ($\Phi_N$), nitridation time ($t_N$) and substrate temperature ($T_S$). In all experiments the impinging N flux was set to $\Phi_N \approx 4.0 \cdot 10^{14}$ atoms/cm²·s, nitridation time was set to $t_N = 10$ min, whereas the substrate temperature was kept same as during the In deposition. The AFM and SEM images of the samples whose fabrication was stopped after the nitridation step, clearly reveal that the resulting InN islands have significantly lower surface density and significantly larger size (compared to the density and size of their metallic In droplets counterparts). In the particular case of the sample fabricated at $T_S \approx 15$ °C, the InN island density was estimated at $\sigma = 4.0 \cdot 10^9$ cm$^{-2}$, their diameters being typically in the range of tens of nanometers (sporadically surpassing the size of 100 nm). It has also been observed that maintaining low substrate temperature (15 °C), during the nitridation step, is crucial to achieve InN QDs with nanometric size (at the end of the fabrication process); the higher substrate temperatures (15 °C < $T_S ≤ 110$ °C) led to lower surface density and progressively bigger InN islands (eventually surpassing 1 μm diameter).

To get further insight into their crystalline properties, the InN islands from Fig. 2b were studied by STEM. The diffraction contrast-TEM micrograph, taken in bright field mode, presented in Fig. 3a resolves three InN islands. A HRTEM micrograph, presented in Fig. 3b reveals further details of their morphology and crystallinity. First, the height and diameter of the inspected island are estimated at $H \approx 2$ nm and $D \approx 46$ nm, respectively. Second, the atomically resolved image reveals that these islands have polycrystalline form, i.e. that they are formed from small monocrystalline grains, and lie on an amorphous-like $SiN_x$ layer (the unambiguous confirmation of chemical composition of surface QDs (InN), amorphous-like interlayer ($SiN_x$) and underlying substrate (Si) is given hereafter). These grains, which show certain degree of misorientation with respect to each other, are separated by domain boundaries, as those indicated with red lines in Fig. 3b. The polycrystallinity is most likely the consequence of low substrate temperature used to convert metallic In droplets into InN; the low substrate temperature ($\approx 15$ °C) prevented crystal re-arrangement into energetically preferential crystal form [20]. Further explanations considering the mechanisms which drive this fabrication step are addressed hereafter.

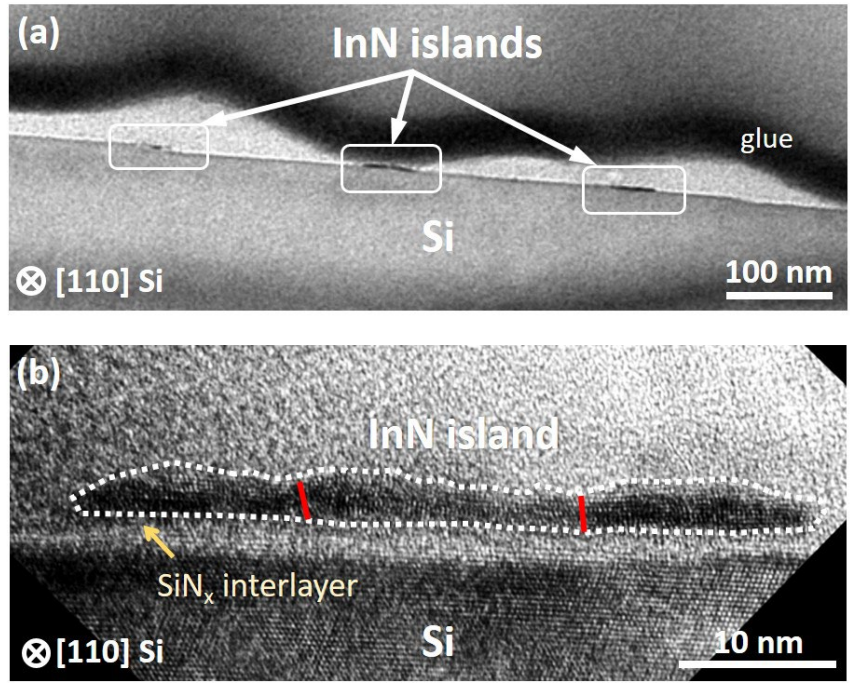

**Fig. 3.** (a) Bright Field DC-TEM micrograph for a cross-section view of extended InN islands on Si(111). (b) HRTEM image of the polycrystalline structure of an InN island grown on an amorphous-like $SiN_x$ layer.

**Stage II: Thermal annealing.** In the final step, InN islands are annealed aiming to allow for their reordering into energetically preferential crystal form and to, consequently, increase their crystalline quality. For that, the substrate temperature is slowly raised (4 °C/min) up to $T_S \approx 300$ °C and then kept constant for $t_{ann} = 30$ min, allowing for InN crystal reordering. After annealing, the sample is cooled down to room temperature. The sample was exposed to active nitrogen during the whole process to prevent possible accumulation of metallic In on the sample surface, which might occur due to undesired InN decomposition (note that InN decomposition in vacuum follows the equation $2InN_{(solid)} \rightarrow 2In_{(liquid)} + N_{2(gas)}$, i.e. the decomposition would result in undesired accumulation of liquid In on the sample surface). SEM and AFM analyses of the samples (Fig. 2d and e) have revealed that InN islands convert into either: (i) a sole (isolated) QD or (ii) a cluster of several QDs. Comparison of morphologies of InN objects before (Fig. 2b) and after (Fig. 2d) annealing, strongly suggests that during annealing step smaller InN islands transform into isolated QDs, while bigger ones transform into clusters of QDs. The QD surface density was estimated at $\sigma \approx 1.5 \cdot 10^{10}$ cm$^{-2}$.

STEM-HAADF image (Fig. 4a) resolves six sole InN QDs. Further detailed HRTEM analysis performed on four of these QDs revealed their average height of $H \approx 6.1$ nm ($\sigma_H = 0.5$ nm) and diameter of $D \approx 9.6$ ($\sigma_D = 0.5$ nm), their average aspect ratio being estimated at $H/D = 0.63 \pm 0.03$ (higher than any previously reported aspect ratio for InN QDs [15,21]). HRTEM image featured in Fig. 4b, shows that QDs have hemispherical-like shape, containing different lateral facets. The contact angle of the QD with the underlying surface is estimated at ≈63°, the corresponding lateral facet thus being identified as $[10\bar{1}1]$ semi-polar one. The total of four different semi-polar QD facets $[10\bar{1}1]$, $[10\bar{1}2]$, $[10\bar{1}3]$ and $[10\bar{1}4]$ have been identified by indexation of their associated diffraction spots in the Fast Fourier Transform (FFT) of the image featured in Fig. 4c [22]. It is worth mentioning that artifacts like the white lines across this figure (see Fig. 4c) appear because FFT calculation was performed over a square-shaped area.

STEM-HAADF images were carried out to unambiguously confirm the chemical composition of surface QDs, amorphous-like interlayer and underlying substrate (Fig. 4d). The intensity in these images is related to the average atomic mass of chemical constituents, as commented in the experimental section [23]. These images were further complemented with EDX maps (Figs. 4e and f). The combination of the two techniques unambiguously confirmed that QDs are composed of In and N solely, that the substrate is composed of Si solely, whereas the amorphous-like, 1-2 nm thick interlayer is composed of Si and N, as expected.

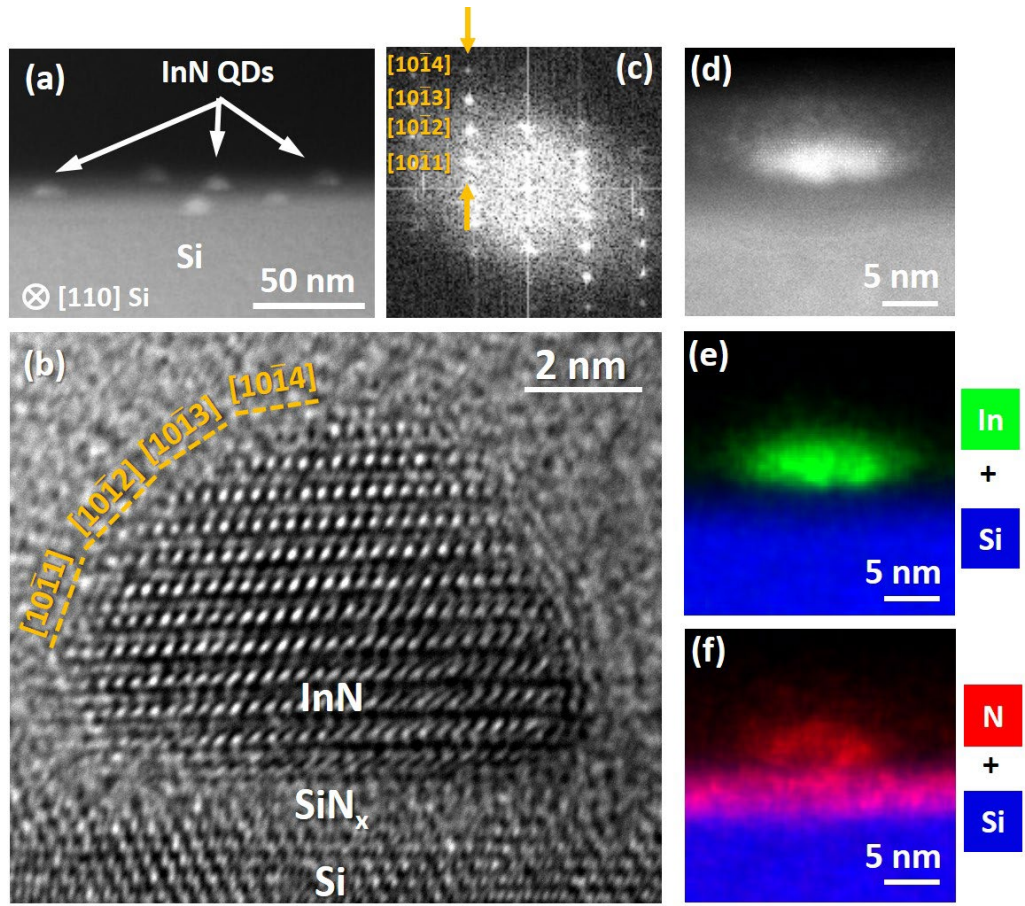

**Fig. 4.** (a) HAADF image showing hemispherical-like shape and high aspect ratio of InN QDs formed across the surface of the Si(111) substrate. (b) HRTEM of a single crystalline wurtzite QD, with lateral facets designated, as guide to the eye. (c) FFT pattern for the InN QD confirming the formation of the $[10\bar{1}1]$, $[10\bar{1}2]$, $[10\bar{1}3]$ and $[10\bar{1}4]$ QD lateral facets (d) HAADF and (e-f) EDX maps of individual InN QD (the color code is: green for In, blue for Si and red for N), unambiguously determining the chemical composition of QDs (InN), amorphous interlayer ($SiN_x$) and substrate (Si).

Making use of HRTEM images, it is possible to study the crystalline structure of individual QDs. Fig. 4b shows the ABAB… atomic ordering expected for a wurtzite (hexagonal) structure when observed along the [11$\bar{2}$0] zone axis. Fig. 4b shows a single QD in which the InN[11$\bar{2}$0] basal direction matches the Si[110] basal direction. It is worth to mention, that coincidence of these two crystal directions (InN[11$\bar{2}$0] ∥ Si[110]) is not very frequent, in other words, the QDs do not show clear "twist-related" epitaxial alignment to the underlying Si(111) substrate. On the other hand, the QDs do show "tilt-related" epitaxial alignment InN[0001] ∥ Si[111] to the underlying substrate.

Before we move to theoretical frame of the exposed fabrication process, we will provide some additional experimental evidence concerning the morphological transition induced by thermal annealing step.

Figure 5 shows histograms analyzing height and diameter of InN islands and InN QDs on Si(111) substrates. Since TEM is an expensive and time-consuming technique, only four QDs were directly assessed by HRTEM. Thus, to make statistical analysis of heights and diameters of InN islands and QDs, AFM and SEM have been used, respectively (since they can easily provide a statistic of bigger samples). Note that while AFM is a good tool to determine height of small surface objects, it is much less trustable to determine diameter of small (and closely packed objects) such as QDs; due to finite radius of the AFM tip (≈ 10 nm), the AFM actually tends to overestimate real diameters. Therefore, SEM has been chosen to estimate diameters. However, at such small scale the SEM is at the very limit of its resolution, thus also prone to errors. Note further that while performing AFM and SEM statistics there is no insight into a crystalline structure of the analyzed objects. Thus, InN island sample probably includes a certain fraction of single, monocrystalline QDs (which achieved their -monocrystallinity before thermal annealing step); similarly, the InN QDs sample probably includes a certain fraction of polycrystalline InN objects (which did not manage to fully transform from polycrystalline islands into a group of monocrystalline objects during the thermal annealing step). In other words, this approach inevitably leads the two statistics to "contaminate" each other.

Bearing in mind that relatively big data samples can be gathered by AFM and SEM, we point out that the shown statistics are a good tool to further quantitatively support the "reorganization scenario", which takes place during the thermal annealing step; both significant increase in height (from 2.0 nm to 5.7 nm) and reduction in diameter (from 61.3 nm to 19.4 nm) are unambiguously observed (Fig. 5). Bearing, however, in mind that SEM is at the very limit of its resolution and that the two statistics partially "contaminate" each other, we also point out that we must be very cautious with the interpretation of mean values obtained via these statistics. Comparing heights and diameters obtained via AFM/HRSEM statistics, with those obtained on the few objects by HRTEM, we actually observe good agreement in InN island/QD height, but overestimation of their diameters (by AFM/SEM).

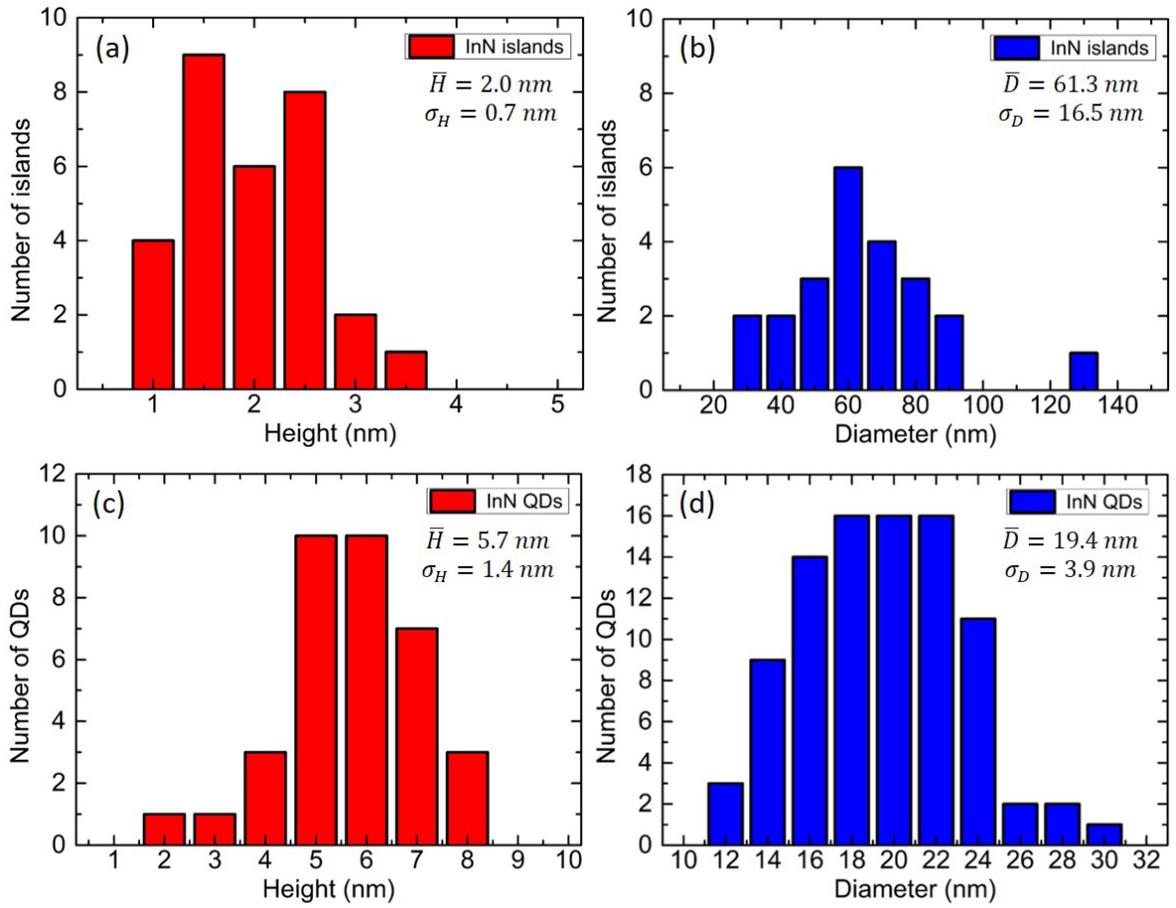

**Fig. 5.** Distribution of surface objects' size on Si (111) substrate. (a) InN islands heights, as estimated by AFM, (b) InN islands diameters, as estimated by SEM, (c) InN QDs heights, as estimated by AFM, and (d) InN QDs diameters, as estimated by SEM. Mean value and standard deviation of each statistic are also shown, for clarity.

### 3.1.2 Theoretical frame

Before we expose a theoretical frame for each of the fabrication steps, we will, for clarity, comment on several terms commonly used when assessing thermodynamic quantities of a system grown by molecular beam epitaxy.

First, we will consider that a system is in a thermodynamic equilibrium when there is no net macroscopic flow of matter or of energy (either within a system or between systems). Second, note that once the state of thermal equilibrium (i.e. state with no macroscopic flow of energy and/or matter) is achieved, the total thermodynamic energy of the system remains constant. Finally, note that the external changes (such as impinging fluxes and/or substrate temperature) may induce changes in kinetics (i.e. more precisely in constituent particles' kinetic energy), and may temporally destabilize the system, driving it finally to another stable state with lower thermodynamic energy [24,25].

**Model for stage I: In deposition.** During this fabrication step, the "cold" ($T_S \approx 15\ °C$), atomically flat Si(111) substrate is exposed to the impinging flux of "hot" In atoms (evaporated at around $T_{In} = 770\ °C$). It was experimentally observed that the substrate temperature variation ($T_S = -10 - 110\ °C$) did not have significant effect on In islands size and density, as long as its temperature was kept below the In melting point ($T_M = 156.6\ °C$). Once it was increased above the In melting point, both a dramatic increase in size and a decrease in density were observed. This is most likely related to the drastic increase of In adatom mobility when substrate temperature surpasses In melting point. Note that during this step, the substrate is intentionally kept cold, to reduce adatoms kinetics, resulting in formation of small and dense metallic droplets.

**Model for stage I: Nitridation.** In the next step, the "cold" substrate with metallic In droplets on the top is now exposed to "hot" atoms of impinging nitrogen flux (note that plasma source has a limited "cracking" efficiency, thus emitting both active atomic N and passive molecular nitrogen $N_2$). On one hand, the exposure of the Si(111) surface to active nitrogen leads to a rapid formation of the thin amorphous-like SiN surface layer, as observed by HRTEM and EDX. On the other hand, the exposure of metallic In droplets to active nitrogen leads to their gradual conversion into semiconductor InN. It has also been clearly observed by SEM and AFM that during/after the nitridation step, the InN islands undergo surface reordering ("agglomeration") into polycrystalline-like plateaus with comparatively bigger size than those of initial In droplets. This process to some extent resembles Ostwald ripening, in the sense that the atoms from the smaller crystals (i.e. sole In droplets or small InN islands) migrate toward and redeposit onto larger crystals (i.e. extended polycrystalline InN islands) [26]. During initial nitridation, some fraction of In converts into InN islands, which then act as nucleation seeds for further In-to-InN conversion. Although the exact mechanism governing this surface reordering remains unknown, it is enabled by enhanced kinetics of the surface species; increased mobility of surface adatoms, In droplets or entire InN islands can be, on one hand, due to (i) additional kinetic energy that these species receive upon their exposure to the hot impinging nitrogen flux and, on the other hand, due to (ii) their weaker interaction with the underlying substrate upon its nitridation (i.e. upon the substrate surface amorphization, from crystalline Si(111) to amorphous $SiN_x$). Once grouped into plateaus-like structures, the InN islands preserve their polycrystalline structure, basically due to "insufficient kinetics" to reorder into another thermodynamically stable state with lower energy.

**Model for stage II.** Combining together and comparing the morphological and structural analyses of polycrystalline InN plateaus with those of final InN QDs, one can observe significant similarity between this conversion process and the one often observed when thin continuous solid films, deposited on foreign substrates, "fracture" into independent 3D surface objects. The latter process is commonly referred to as thin film solid state dewetting [27]. The theory predicts that if a polycrystalline island is given enough thermal energy and time it may undergo reordering, breaking up at the "energetically costly" grain boundaries (sometimes completely eliminating them) and transforming into a cluster of separate (mono)crystalline objects (bearing in mind that we referred to the surface objects gathering together as "agglomeration", we will refer now to their separation as "deagglomeration") [28].

Note further that in stage I (during both In deposition and nitridation) there is a flow of both matter and energy, from the environment toward the studied epitaxial film - the epitaxial film both grows and reorders. During stage II, the flow of matter is internal, exclusively, i.e. the epitaxial film only reorders. In stage II, the kinetic energy of the surface species is enhanced thermally, allowing them to reorder into a new thermodynamically stable system with, in total, lower energy. In that sense, unlike the evolution of the epitaxial film during stage I (when there is a significant "flow of matter" from the environment toward the system), the evolution of the epitaxial film during stage II can be easily linked to the evolution of distinct thermodynamic quantities in the system (since the "flow of matter" in this case is exclusively internal). The final result of this reordering is a new stable system with reduced total energy. Below we give a brief qualitative analysis of the thermodynamic quantities, tightly linked to the epitaxial film reordering, as suggested by Shchukin and Bimberg [24,25]. We point out that are goal is not a deep quantitative analysis of these quantities, but rather their qualitative linking to the observed surface reordering of the epitaxial film, for improved clarity.

In the particular case of polycrystalline InN plateaus, they are composed of atoms which, as a function of their energy, can be roughly divided into four different groups [24,25,29,30]:

(i) grain boundary atoms, located at or close to the boundaries between different monocrystalline domains;

(ii) surface atoms, located at or close to the free crystal facets;

(iii) interface atoms, located at or close to the underlying interface with substrate;

(iv) bulk atoms, located inside the monocrystalline domains.

If the total energy per unit volume of an infinite, defect- and strain-free InN crystal is taken as a reference ($E_{bulk} = 0$), the energy of atoms may be locally altered due to their proximity to grain boundaries, free crystal surface or interface or due to the (local) presence of strain. The total free-energy per unit volume of a polycrystalline plateau thus can be written as:

$$E_{\text{total polycrystalline I}} = E_{\text{grain boundaries I}} + E_{\text{surface I}} + E_{\text{interface I}} + E_{\text{strain I}}, \quad (1)$$

where sub-index I stands for "stage I".

After thermal annealing, a polycrystalline plateau reorders into a group of monocrystalline QDs, as observed by TEM. The thermal annealing contributes to the elimination of grain boundaries. Consequently, concerning monocrystalline InN QDs, they are composed of atoms which as a function of their energy can be roughly divided into three different groups (surface, interface and bulk atoms). The total energy-per-unit-volume of a monocrystalline QD can thus be written as:

$$E_{\text{total monocrystalline II}} = E_{\text{surface II}} + E_{\text{interface II}} + E_{\text{strain II}}, \quad (2)$$

where sub-index II stands for "stage II".

The difference in total energy-per-unit-volume of the two distinct morphologies, is likely the driving mechanism for the observed shape transition, that is: $E_{\text{total monocrystalline II}} < E_{\text{total polycrystalline I}}$. The reduction in total energy, due to the morphological transition is:

$$\Delta E_{\text{total}} = E_{\text{total monocrystalline II}} - E_{\text{total polycrystalline I}} = \Delta E_{\text{grain boundaries}} + \Delta E_{\text{surface}} + \Delta E_{\text{interface}} + \Delta E_{\text{strain}} < 0. \quad (3)$$

The energetically dominant term on the right hand side of Eq. 3 will drive the morphological transition.

By comparing the morphology of polycrystalline plateaus with the morphology of groups of single QDs (AFM and SEM images shown in Fig. 2), we can get an insight into the sign of the thermodynamic quantities on the right-hand-side of Eq. 3. First, the reordering eliminates the grain boundaries, and supposing that their elimination is complete, it yields $\Delta E_{\text{grain boundaries}} = - E_{\text{grain boundaries I}} < 0$. Second, the elimination of the grain boundaries leads to a reordering of the free crystal surface. The total area of the free crystal surface now increases significantly, yielding $\Delta E_{\text{surface}} = E_{\text{surface II}} - E_{\text{surface I}} > 0$. Third, inspecting the AFM and HRTEM images, before and after the morphological transition, one can observe that the change in the interface area (between InN and SiN) is small, and, in particular, much smaller than the change in the area of the free crystal surface of the two distinct morphologies. Bearing further in mind that the density of energy stored at "crystal – amorphous interface" is significantly lower than the density of energy stored at free crystal surfaces [29], we conclude that the term related to the interface energy change can be neglected with respect to the term related to the crystal surface energy change, that is: $|\Delta E_{\text{interface}}| \ll |\Delta E_{\text{surface}}|$. Finally, the elimination of grain boundaries combined with the increase in total crystal surface both favor the reduction of strain, that is: $\Delta E_{\text{strain}} = E_{\text{strain II}} - E_{\text{strain I}} < 0$.

Bearing the previous analysis in mind, Eq. 3 simplifies into:

$$\Delta E_{\text{total}} \approx - |E_{\text{grain boundaries I}}| + |\Delta E_{\text{surface}}| - |\Delta E_{\text{strain}}| < 0. \quad (4)$$

We identify energy reduction due to elimination of grain boundaries ($E_{\text{grain boundaries}} \downarrow$) as the driving mechanism for the observed. Next, the elimination of grain boundaries leads to crystal reordering, resulting in increase in free crystal surface and, consequently, in increase of its corresponding energy term ($E_{surface} \uparrow$). Finally, as a consequence of both grain boundary elimination and free crystal surface increase, the strain, and its corresponding energy term, are likely reduced ($E_{strain} \downarrow$).

To conclude this section, we will comment on the issue related to QDs shape stability. To estimate if the stable QDs shape has been reached (i.e. if a new "thermodynamic equilibrium" has been reached) when Stranski-Krastanov method is used, the system is commonly monitored *in-situ* by reflective high energy electron diffraction (RHEED) [31]. To induce Stranski-Krastanov transition, the system is driven out of the equilibrium normally by cutting the impinging fluxes (i.e. by reducing the overpressure above the thin epitaxial layer); then sufficient time is given to the system (under these new kinetics conditions) the surface species to reorder into a stable form. This reordering usually takes few seconds or minutes. After this transient time, a new stable state with reduced total energy is reached [32].

In our case, due to very low temperatures used for the QD fabrication, very small amount of surface material and a very weak epitaxial relationship between the surface species and the underlying substrate, the RHEED *in-situ* monitoring was not possible. Thus, in the final nitridation step, the sample was given 30 minutes to reach new stable state. Note that the given time is roughly one order of magnitude longer than typical reported transient times for Stranski-Krastanov transition [31], In addition, an indirect proof that the stable QDs shape is finally reached is that virtually identical surface object shapes were observed across entire surface area and on many (several tens of) analyzed samples.

## 3.2. Influence of substrate on InN QDs formation: Si(111) vs. Si(001) and In$_{0.3}$Ga$_{0.7}$N/Si(111)

To check for the validity of the proposed growth model (and the overall potential of the presented fabrication method) the DE epitaxy was tested on two more substrates: atomically flat Si(001) and relatively rough In$_{0.3}$Ga$_{0.7}$N/Si(111) templates. Concerning the selection of the two substrates, note that when switching from Si(111) to Si(001) substrate, the substrate surface remains exceptionally (atomically) flat, however the apparent arrangement of surface Si atoms changes significantly, from 3-fold symmetry ("hexagonal-like"), in the former, to 4-fold symmetry (cubic), in the latter case. On the other hand, when switching from Si(111) to In$_{0.3}$Ga$_{0.7}$N/Si(111) substrate, the substrate surface changes from atomically flat in the former, to relatively rough one (rms ≈ 10 nm) in the latter case. However, the apparent arrangement of the surface atoms does not change very significantly; it changes from 3-fold symmetry ("hexagonal-like"), in the former, to the 6-fold symmetry (hexagonal), in the latter case. In addition, both Si(111) and Si(001) surfaces are prone to surface nitridation (as confirmed hereafter), i.e. to the formation of the amorphous SiN$_x$ surface layer. Consequently, the InN QDs are hosted on amorphous-like (SiN$_x$) layer, in the case of Si(111) and Si(001) substrates, and on crystalline (In$_{0.3}$Ga$_{0.7}$N) layer, in the case of In$_{0.3}$Ga$_{0.7}$N/Si(111) template. These particular differences between the three substrates and their consequent influence on the QD fabrication are addressed in this section.

### 3.2.1 InN QDs on Si(001)

In order to perform a comparative study, the growth conditions used for InN QD fabrication on Si(001) were identical to the previously described ones, i.e. to those used for their fabrication on Si(111). To gather experimental evidence, the fabrication procedure was stopped after the first and the second stage (i.e. before and after thermal annealing, respectively). As expected, results very similar to those obtained on Si(111) were obtained, as shown below.

AFM and SEM investigation of InN islands formed after stage I on Si(001) surface (Figs. 6a and b), revealed that their shape and size distribution are similar to those fabricated on Si(111). Note that despite cubic (i.e. 4-fold symmetry) arrangement of surface Si(001) atoms, AFM scans of individual InN islands (Fig. 6b) show that these islands have a "close-to-hexagonal" shape (though usually not clearly defined), suggesting that InN atoms are internally arranged in the form of hexagonal wurtzite crystal.

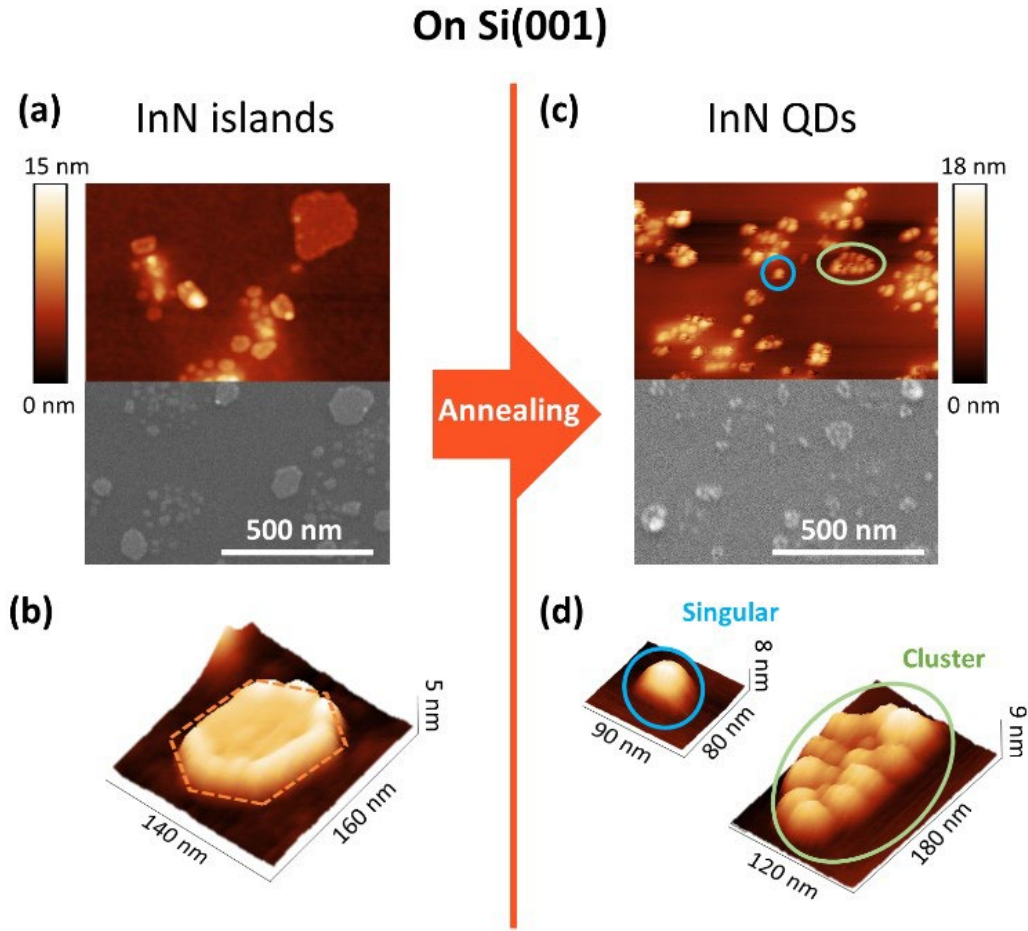

**Fig. 6.** Change of shape of InN nanostructures on Si(001) substrate from (a,b) islands to (c, d) QDs upon annealing as revealed by AFM and SEM measurements.

Annealing of InN islands on Si(001) resulted in their reshaping into InN QDs. Again, single QDs and clusters of QDs can be distinguished (Figs. 6c and d), the formation scenario being very similar to the Si(111) case, smaller InN islands transforming into isolated QDs and bigger InN islands transforming into clusters of QDs. Making use of AFM measurements, surface density of InN QDs was estimated at $\sigma \approx 1.6 \cdot 10^{10}$ cm$^{-2}$.

HRTEM structural analysis of InN islands grown on Si(001) (Fig. 7a), confirmed their polycrystalline structure with twisted domains and no clear alignment to Si[001] crystal direction. In the few cases where it was possible to measure interplanar distances, the following values were extracted: $d_{11\bar{2}0}$=1.77 Å, corresponding to the interplanar wurtzite distance $d_{hkil}$ for the (11$\bar{2}$0) planes, and $d_{0002}$=2.77 Å, corresponding to the interplanar wurtzite distance of the (0002) ones. These results are in a good agreement with calculated values for the fully relaxed wurtzite structure ($d_{11\bar{2}0}$=1.766 Å and $d_{0002}$=2.846 Å) [33]. The good agreement between measured and expected crystal plane distances supports previously established hypothesis, that InN nanostructures on Si(001) are, indeed, formed from hexagonal (wurtzite) crystal domains.

On the other hand, HRTEM images of InN QDs revealed their single-crystalline structure, like in the case of QDs grown on Si(111). In most cases the atomic fringes in the HRTEM images are parallel (or almost parallel) to the substrate surface (Fig. 7b), whereas in some cases the fringes are tilted from 30° to 45° (Fig. 7c). The fact that only atomic fringes (and not columns) were resolved, suggests that QDs are randomly rotated along the vertical (001) Si(001) crystal direction. The measured interplanar distances within the QDs were $d_{11\bar{2}0}$=1.71 Å, and $d_{0002}$=2.85 Å, which, when compared with the calculated values for a fully relaxed InN hexagonal wurtzite crystal, as previously indicated, confirm that QDs crystalize in the hexagonal wurtzite form, despite being grown on cubic Si(001) substrate. Finally, detailed HRTEM analysis of two selected single QDs found their heights and diameters to be very similar H ≈ 5.7 nm and D ≈ 8.7, their aspect ratio being estimated at H/D ≈ 0.65. Note that QDs grown on Si(111) and Si(001) have very similar density, shape and size, as expected.

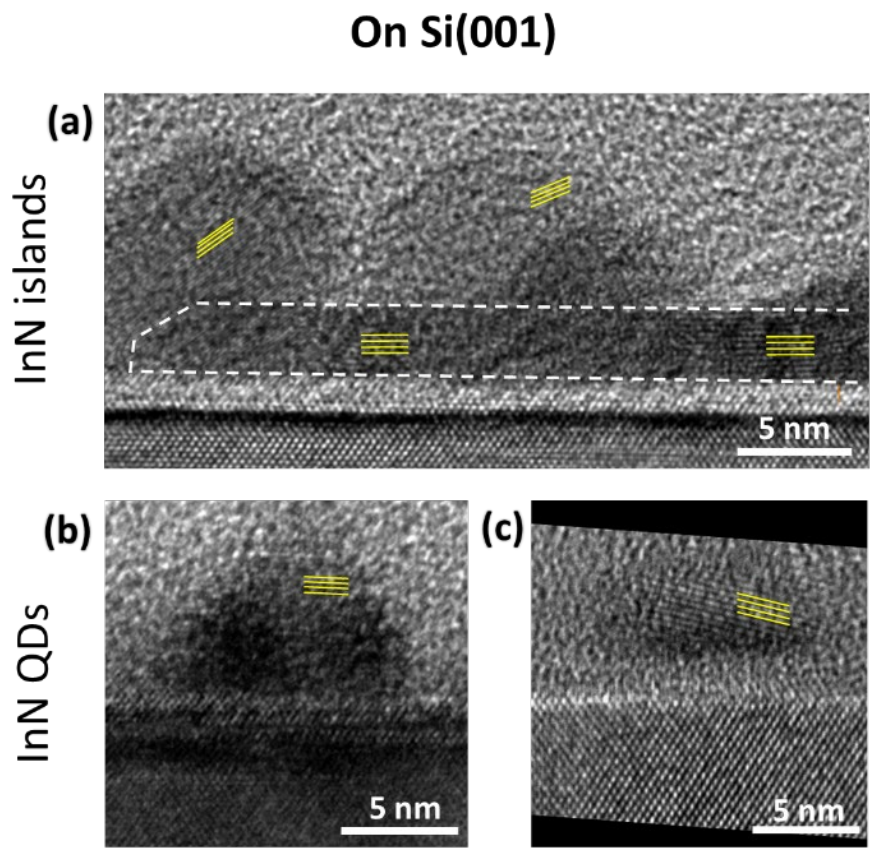

**Fig. 7.** HRTEM micrographs of InN (a) islands and (b) and (c) QDs, grown on Si(001) substrate. Yellow lines indicate atomic fringes.

Finally, in Fig. 8 we show distribution of heights and diameters of the InN islands and QDs, on Si(001) substrate, for completeness. The same experimental procedure like in the case of their "on-Si(111)-counterparts" has been applied, for consistency. The same trend of increase in height (from 2.8 to 5.4 nm) and decrease in diameter (from 29.4 to 24.0 nm) has been observed, suggesting that the thermal annealing had very similar impact on surface reordering like in the previously described case, as expected.

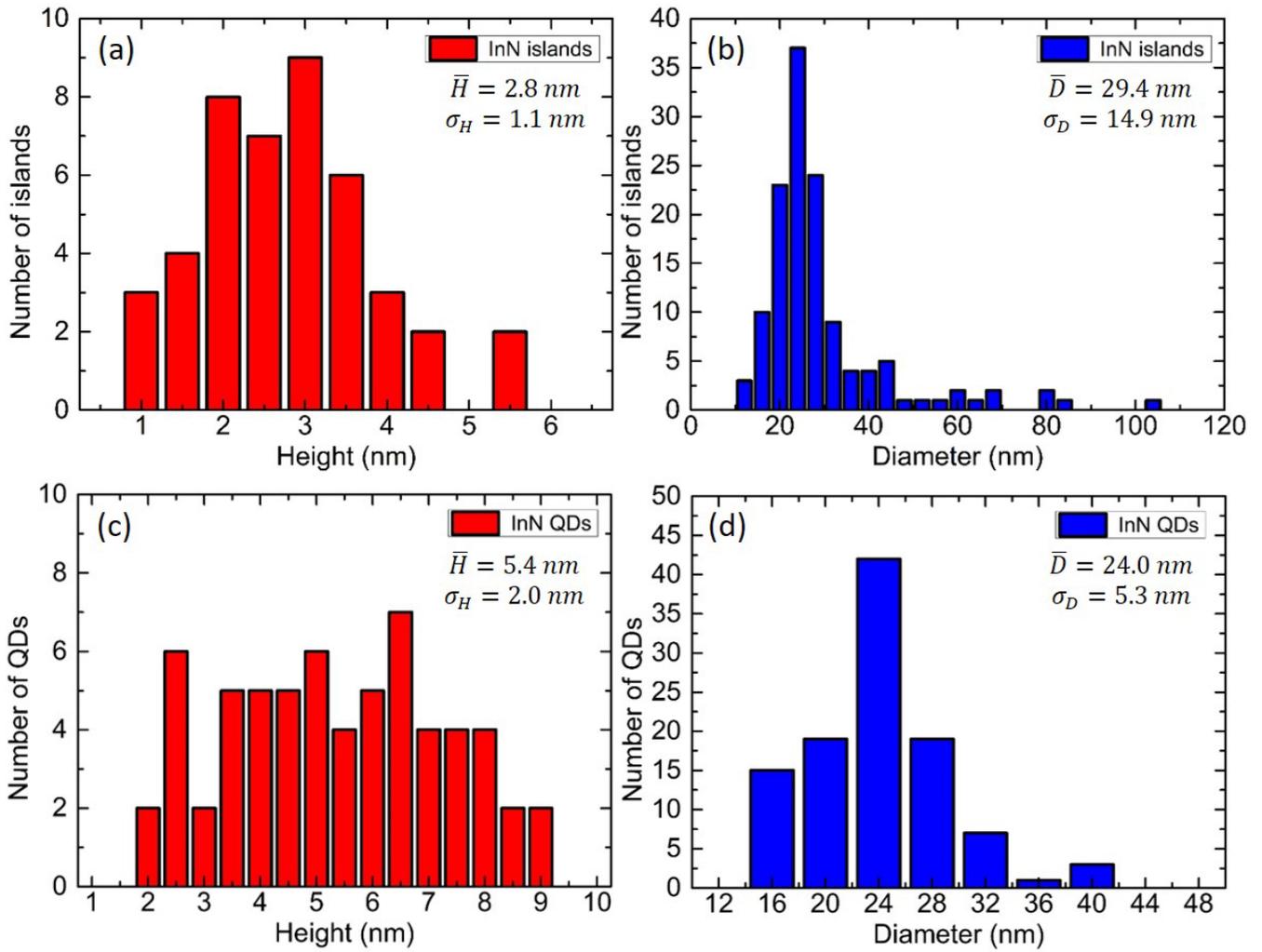

**Fig. 8.** Distribution of surface objects' size on Si(001) substrate. (a) InN islands heights, as estimated by AFM, (b) InN islands diameters, as estimated by SEM, (c) InN QDs heights, as estimated by AFM, and (b) InN QDs diameters, as estimated by SEM. Mean value and standard deviation of each statistic are also shown, for clarity.

### 3.2.2 InN QDs on In$_{0.3}$Ga$_{0.7}$N/Si(111)

Finally, InN QDs fabrication procedure was applied to the most challenging case, rough (*rms* ≈ 10 nm) In$_{0.3}$Ga$_{0.7}$N/Si(111) substrate. The latter procedure also resulted in fabrication of single crystalline wurtzite InN QDs, with only few variations from the previous Si substrate cases, as detailed below.

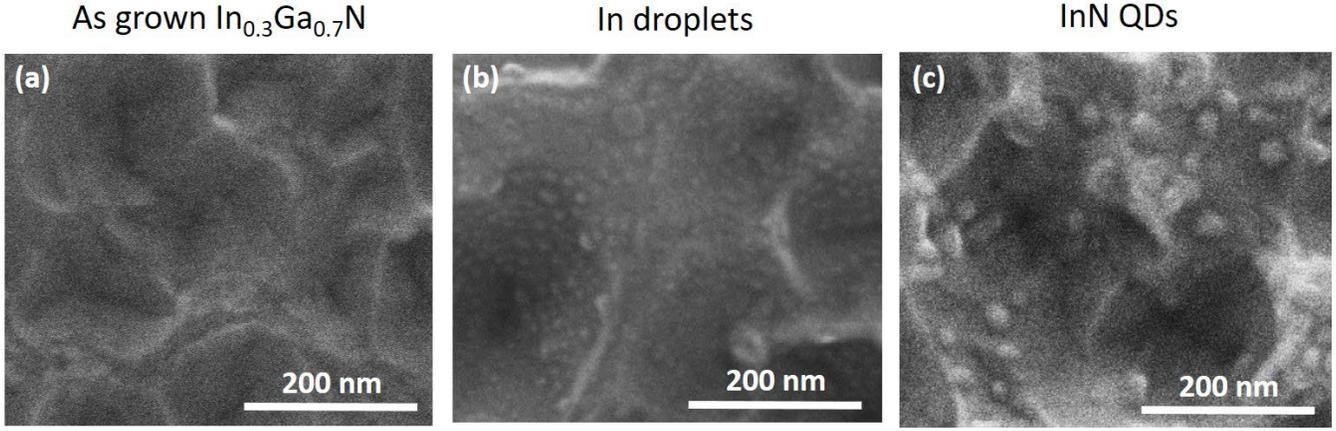

**Fig. 9.** SEM images of planar views of the surface of $In_{0.3}Ga_{0.7}N$ for the main steps during the InN QDs fabrication on $In_{0.3}Ga_{0.7}N/Si(111)$: (a) as grown $In_{0.3}Ga_{0.7}N$, (b) with deposited In droplets and (c) with final InN QDs.

In order to facilitate characterization of InN QDs they were grown under identical growth conditions as for Si(111) and Si(001) cases, apart from the deposition time for metallic In droplets ($t_{In}$) which was increased from 10 to 15 seconds. The longer deposition time was set to obtain somewhat larger QDs and, consequently, facilitate their analysis on the rough $In_{0.3}Ga_{0.7}N$ surface. The formation stages of InN QDs on the $In_{0.3}Ga_{0.7}N/Si(111)$ template were examined *ex situ* by SEM, as shown in Fig. 9. The "as grown" $In_{0.3}Ga_{0.7}N$ layer, shown in Fig. 9a, exhibits a much rougher surface (rms ≈ 10 nm) compared to the Si substrates used before. That high roughness impeded analysis by AFM. The surface roughness also imposes a challenge to fabricate QDs on the surface top, since standard Stranski-Kratanov technique requires flat plateaus for 2D-to-3D morphological transition and is, thus, unlikely to "work" properly on such rough surfaces [10,34].

Upon In metal deposition, closely packed droplets were formed with a density of $\sigma \approx 2.6 \cdot 10^{10}$ cm$^{-2}$ and an average diameter of $D \approx 13$ nm as shown in Fig. 9b, similarly to Si case. However, the InN QDs, formed at the end of stage II (i.e. after nitridation and thermal annealing, Fig. 9c) were found to have several significant differences, compared with their "on-Si-counterparts". First, no clusters of QDs were found in this case. Now the QD ensembles consist of sole objects, with a relatively high and uniform density ($\sigma \approx 7.4 \cdot 10^{9}$ cm$^{-2}$) over the sample surface. We attribute this difference to the fact that during the nitridation step the In droplets and/or sole InN islands now reorder on crystalline $In_{0.3}Ga_{0.7}N$ surface (vs. amorphous $SiN_x$, in the previous cases), having thus probably lower surface mobility. In addition, these QDs have a somewhat larger size compared to their "on-Si-counterparts". Making use of few HRTEM images, their average height and diameter were estimated at $H \approx 11$ nm and $D \approx 30$ nm (Figs. 9c, 10a and 10b). Finally, the average InN QDs aspect ratio was estimated at $H/D \approx 0.37$, significantly lower than in the case of their "on-Si-counterparts".

A structural study of the QDs was realized via combination of conventional STEM and HRTEM. In the HRTEM image of the InN QD in Fig. 10b, clear atomic columns (not only fringes) are identified in both $In_{0.3}Ga_{0.7}N$ and InN regions, suggesting that there are both in-plane (InN[0001] || $In_{0.3}Ga_{0.7}N$[0001]) and vertical (InN[10$\bar{1}$0] || $In_{0.3}Ga_{0.7}N$[10$\bar{1}$0]) epitaxial relations between the InN and $In_{0.3}Ga_{0.7}N$. The interface between InN and $In_{0.3}Ga_{0.7}N$ is clearly heteroepitaxial, as expected (Fig. 10b). InN[0001] direction is found slightly titled (≈ 1°) in respect to the $In_{0.3}Ga_{0.7}N$[0001]. The InN interplanar atomic distances were estimated at $d_{11\bar{2}0}$=1.758 Å and $d_{0002}$=2.868 Å, in a good agreement with the calculated theoretical values for relaxed wurtzite InN. The residual strain in the QD has been traced via differences between the measured interplanar InN distances and those theoretically calculated, for fully relaxed InN crystal. The obtained results indicate high degree of intra-dot strain relaxation, which is likely facilitated via large free surface of the QD itself [33].

Angles between InN QD sidewall facet and the In$_{0.3}$Ga$_{0.7}$N(0001) plane were typically measured to be between 32° and 35°, allowing their assignment to the (10$\bar{1}$3) InN crystal planes (the exact calculated angle in the case of fully relaxed wurtzite InN is 32.1°). Note that this type of faceting is the most frequently reported one for III-Nitride QDs [35].

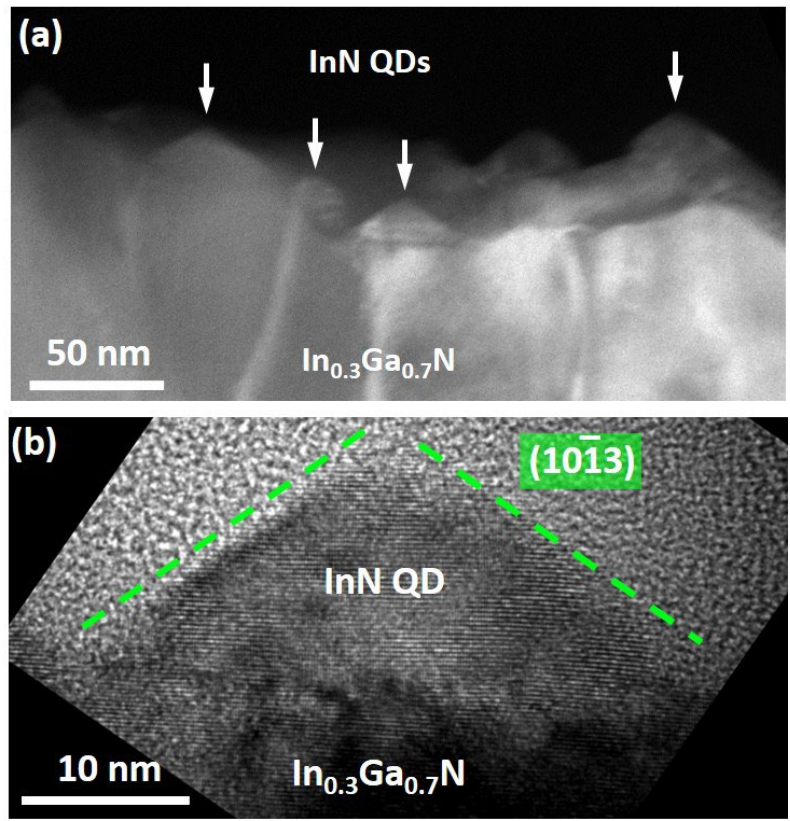

**Fig. 10.** (a) HAADF and (b) HRTEM images of InN QDs grown on an In$_{0.3}$Ga$_{0.7}$N/Si template. InN QD facets are indicated in the HRTEM micrograph.

Finally, in Fig. 11 we show distribution of diameters of the QDs, on In$_{0.3}$Ga$_{0.7}$N/Si(111) substrate, obtained via analysis of top-view SEM images, for completeness (the estimation of height via AFM was not possible due to high surface roughness of the underlying template). The mean value of QDs diameter of 28.1 nm was found and it is in a good agreement with the average diameter found from the analysis of few HRTEM images.

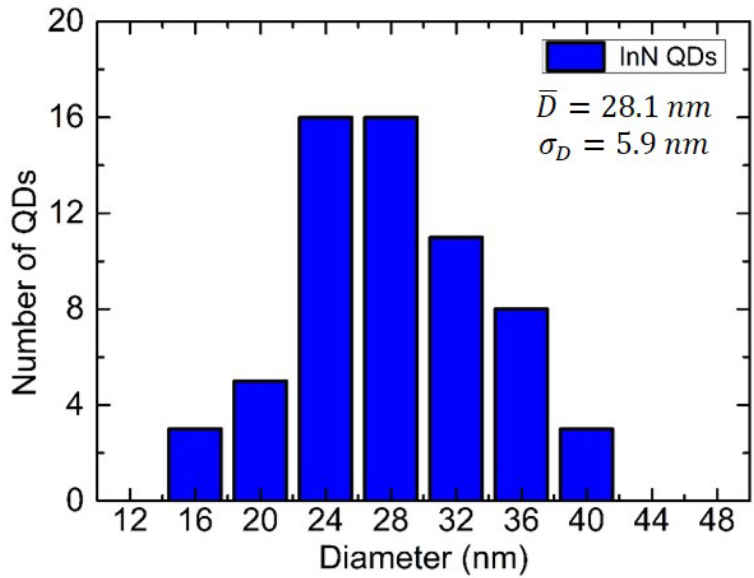

**Fig. 11.** Distribution of InN QDs diameters, as estimated by HRSEM, on In$_{0.3}$Ga$_{0.7}$N/Si(111) templates. Mean value and standard deviation of the statistics are also shown, for clarity.

Concerning the morphological properties of final InN QDs, we observe that they are strongly influenced by the underlying substrate. Namely, the "contact angles" of ≈63° versus ≈32° were observed for QDs fabricated on amorphous-like SiN$_x$ versus crystalline In$_{0.3}$Ga$_{0.7}$N, respectively. This difference is a direct consequence of the type of semipolar contact sidewall facets, found in the former $(10\bar{1}1)$ vs. latter $(10\bar{1}3)$ case, and it has been attributed to a weaker vs. stronger InN interaction with the underlying substrate, respectively. Namely, in the latter InN – In$_{0.3}$Ga$_{0.7}$N case, the stronger crystal-to-crystal interaction (vs. weaker crystal-to-amorphous InN-SiN$_x$ interaction) favors formation of more extended (InN/In$_{0.3}$Ga$_{0.7}$N) hetero-interface. Note that this, further, leads to a lower aspect ratio of InN QDs fabricated on crystalline In$_{0.3}$Ga$_{0.7}$N (≈0.37) vs. amorphous SiN$_x$ (typically >0.60) surface. Similar aspect ratios have been observed in the case of other III-nitride QDs grown on crystalline substrates (such as GaN/AlN QDs formed via Stranski-Krastanov [35]) or in the initial phase of the self-assembled GaN nanocolumn nucleation on crystalline AlN templates [36]. In all these cases, pyramidal (sometimes truncated pyramidal) objects, epitaxially aligned to the underlying crystalline substrate were reported, the lateral semi-polar facet being identified as [$10\bar{1}3$]. The similarity between these crystal systems is most likely due to: (i) combination of III-nitride materials, which have relatively similar properties and (ii) the strong "crystal-to-crystal" interaction, between surface nanobjects and the hosting underlying substrates, which is present in all mentioned cases, and which seems to have a strong impact on the morphology of nanocrystals formed on the substrate surface, "preferring" formation of extended hetero-interface between the two.

Concerning the crystalline properties of InN QDs, the QDs are found to be single crystal hexagonal wurtzite objects, in all three studied cases. Note that they are found hexagonal even when cubic Si(001) was used as an underlying substrate. This result, however, is not surprising since in this case the formation on InN QDs takes place on amorphous-like SiN$_x$, thus it is not (significantly) influenced by the underlying Si(001) crystalline structure. Once again, the results concerning InN nanostructures crystallinity are found in qualitative agreement with a wide variety of other III-nitride nanostructures grown on crystalline (AlN, GaN, Al$_x$Ga$_{1-x}$N) and amorphous-like (SiN$_x$) layers. In particular, we emphasize that self-assembled GaN nanocolumns grown on Si(001) also have hexagonal wurtzite crystal structure [37].

# 4. Conclusion

In this work, it is shown that droplet epitaxy allows the fabrication of single crystalline InN quantum dots on atomically flat Si ((001) and (111)) as well as on relatively rough (rms ≈ 10 nm) In$_{0.3}$Ga$_{0.7}$N/Si(111) templates. The applied procedure consists of two fabrication stages. During the first stage, the cold substrate ($T \approx 15$ °C) is first exposed to an impinging In flux, resulting in the formation of solid In droplets on the substrate surface, and then to an impinging active nitrogen flux, resulting in In droplets crystallization and a consequent formation of solid polycrystalline InN islands. During the second stage, the substrate, which is still kept exposed to active nitrogen, is heated up to $T \approx 300$ °C, to allow for polycrystalline InN islands reordering into groups of independent single-crystalline wurtzite InN QDs. Making use of three different microscopies (SEM, AFM and (S)TEM), a detailed experimental insight into the two fabrication stages was achieved. The two stages are addressed (and qualitatively explained) within the scopes of adatom surface kinetics (stage I) and total-energy-per-unit-crystal-volume minimization (stage II) (i.e. within the same theoretical frame used to explain formation mechanisms of, much more studied, self-assembled and selective-area-grown GaN nanowires [24,29,30]). Finally, the formation mechanisms of these particular InN QD nanostructures on three different substrates (Si(111), Si(001) and In$_{0.3}$Ga$_{0.7}$N/Si(111)) have also been explicitly compared and linked to the formation mechanisms of other well-known III-N nanostructures, such as self-assembled GaN/AlN quantum dots or self-assembled and selective-area-grown GaN nanowires. The understanding of the driving mechanisms for the formation of this nanostructure opens the door for the new approaches to their employment in III-nitride semiconductor devices.

## Acknowledgments

This work was partially supported by Spanish MINECO Research Grant Number MAT2015-65120-R.